\documentclass[journal]{IEEEtran}
\usepackage{xcolor}
\usepackage{graphicx}
\usepackage{algpseudocode}
\usepackage{amsmath}
\usepackage{booktabs}
\usepackage{multirow}
\usepackage{balance}

\usepackage[ruled,vlined]{algorithm2e}
\ifCLASSINFOpdf
\else
\fi
\begin{document}


\title{On Scalable Design for User-Centric Multi-Modal Shared E-Mobility Systems using MILP and Modified Dijkstra’s Algorithm}
%
%
%
\author{Maqsood Hussain Shah, Ji Li \textit{Member, IEEE} and Mingming Liu \textit{Senior Member, IEEE}
\thanks{M H Shah, and M Liu are with the Insight Centre of Data Analytics and the School
 of Electronic Engineering, Dublin City University, Dublin, Ireland. J. Li is with the Department of Mechanical Engineering, University of Birmingham, UK. \textit{Corresponding author: Mingming Liu. Email: {\tt mingming.liu@dcu.ie}.}}}

\maketitle

\begin{abstract}
In the rapidly evolving landscape of urban transportation, shared e-mobility services have emerged as a sustainable solution to meet growing demand for flexible, eco-friendly travel. However, the existing literature lacks a comprehensive multi-modal optimization framework with focus on user preferences and real-world constraints. This paper presents a multi-modal optimization framework for shared e-mobility, with a particular focus on e-mobility hubs (e-hubs) with micromobility. We propose and evaluate two approaches: a mixed-integer linear programming (MILP) solution, complemented by a heuristic graph reduction technique to manage computational complexity in scenarios with limited e-hubs, achieving a computational advantage of 93\%, 72\%, and 47\%  for 20, 50, and 100 e-hubs, respectively. Additionally, the modified Dijkstra’s algorithm offers a more scalable, real-time alternative for larger e-hub networks, with median execution times consistently around 53 ms, regardless of the number of e-hubs. Thorough experimental evaluation on real-world map and simulated traffic data of Dublin City Centre reveals that both methods seamlessly adapt to practical considerations and constraints such as multi-modality, user-preferences and state of charge for different e-mobility tools. While MILP offers greater flexibility for incorporating additional objectives and constraints, the modified Dijkstra's algorithm is better suited for large-scale, real-time applications due to its computational efficiency.
\end{abstract}

\begin{IEEEkeywords}
Shared E-mobility, MILP, Scalability Analysis, Dijkstra's Algorithm, Multi-modal Traffic Optimization 
\end{IEEEkeywords}

%
\IEEEpeerreviewmaketitle

\section{Introduction}

    \IEEEPARstart{C}{limate} change has become one of the most daunting hallenges of recent times. 
    Recent data from the Environmental Protection Agency (EPA) shows that the transportation sector accounts for about a quarter of greenhouse gas emissions \cite{SOLAYMANI2019989}. Concerns about the environment, urban congestion, and the demand for affordable travel have driven a shift towards alternative transportation modes which has been facilitated by advances in technology and new ideas, such as the integration of electric vehicles, improved public transit, and ride-sharing services \cite{useofsharedemobility}. However, there is still a huge potential for improvement in these emerging alternate ways of travelling \cite{Survey1}. Not only do we need an increased consciousness about the ecological impact of traditional transportation methods, but we also need to ensure means to boost the user confidence to shift towards eco-friendly and energy-efficient options \cite{yan2023review}. 
    Integration of micro e-mobility solutions in the transportation landscape is happening at a gradual pace, however, to make it an essential part of the peoples lives and enhance its usage over a broad range will require further efforts. Range anxiety coupled with lack of user-centric recommendation system is one of the factors when it comes to deter the user-confidence in e-mobility \cite{yan2024u}. Additionally, current research in this domain demonstrates a limited focus on open-source contributions \cite{shah2024optimal}.
    
    Existing works relevant to this domain have tackled a broad range of aspects. For instance, \cite{Deeppool} provides a ridesharing optimization based on deep Q-network (DQN), \cite{DeploymentOpt} provides an optimization of the deployment strategy for shared e-mobility infrastructure. In \cite{DynamicRoutingforSharedemobility}, the alternating minimizing algorithm is used for sorting the routing problem. In \cite{PedistrianDRL}, authors have presented deep reinforcement learning (DRL) based route planning. Similarly, in \cite{decentralizedPath}, authors use DRL based path planning algorithm in the context of automated guided vehicles. In \cite{multimodal_parking}, authors propose a multi-modal path planning model with primary emphasis on finding the optimal parking location. In a different context, the authors in \cite{FERRARA2019409} provide a simulation model for the choice behaviors for e-vehicles with main emphasis on privately owned e-vehicles and transition with the public transport.  More comprehensive survey for similar problems has been presented in \cite{SurveyCompleteOpt}.
    In summary, the existing body of work on route planning in transportation systems, including ride-sharing, car-pooling, and dial-a-ride, offers valuable insights and solutions. However, there remain significant gaps that have yet to be fully addressed in the literature. Specifically, two critical aspects are notably underexplored: the development of a comprehensive, open-source, multi-modal, and user-centric optimization framework, and the challenge of ensuring its scalability and thorough evaluation in real-world scenarios.
    
    While the study in \cite{multimodalenergyefficient} makes strides toward addressing these issues by proposing a multi-modal routing framework that incorporates car and bike sharing, it falls short in several key areas. Notably, the lack of detailed methodology and the absence of an open-source implementation significantly limit the reproducibility and broader applicability of its findings. This highlights the pressing need for more robust frameworks that are not only accessible and transparent but also capable of scaling effectively across diverse urban environments. The work presented in this paper seeks to fill these gaps by offering a solution that is both comprehensive and rigorously evaluated.

This work is part of the broader open-source platform for shared e-mobility which provides a comprehensive emulation platform for generating diverse traffic patterns and deployment scenarios in the context of shared e-mobility. The current work addresses two key aspects of optimization in shared e-mobility. Firstly, we provide a concise formulation of the problem with constrained multi-modal optimization objectives and employ open-source and commercial MILP solvers and modified Dijikstra's algorithm. We particularly focus on the seamless integration and flow of micro e-mobility options (e-bikes and e-scooters) within an exclusive problem setting and considering realistic urban dynamics to provide a user-centric and multi-modal optimization framework. Secondly, we discuss and evaluate both methods from scalability and performance perspective which includes a novel naive graph reduction method to make the MILP based solution more scalable. 
The primary purpose of this paper is to timely report some modest findings that we believe are of essence for the researchers working in the domain of shared e-mobility optimization domain. 
The codes and data required to reproduce the results which could be freely used for further research are provided through the project's github repository\footnote{https://github.com/SFIEssential/Essential}.  
The structure and main contributions in this paper are summarized in the following:
\begin{itemize}
\item In Section II, we present a detailed problem illustration and formulation for multi-modal, user-centric optimization framework amid practical constraints with particular focus on e-hub based deployment with micro e-mobility.

\item Section III introduces two optimization approaches: an MILP and a modified Dijikstra's algorithm both tailored for shared e-mobility. In addition, to address the computational complexity in MILP, we propose a heuristic graph reduction technique that improves computational efficiency for scenarios with limited e-hubs and discuss its limitations. 

\item In Section IV, we conduct a thorough evaluation of the proposed methods, focusing on optimality, scalability, and adaptability to multi-modal scenarios and constraints. We evaluate various MILP solvers (Gurobi \cite{gurobi}, PuLP (CBC)\cite{PulpCBC}, GLPK \cite{GLPK}) alongside the modified Dijkstra's algorithm. The evaluation utilizes real-world traffic patterns and deployment scenarios generated using the platform \cite{shah2024optimal}, based on the Dublin City Centre map. Finally, Section V provides the concluding remarks.
\end{itemize}
\begin{figure}[htb]
\centering
\includegraphics[width=0.5\textwidth, height = 6cm]{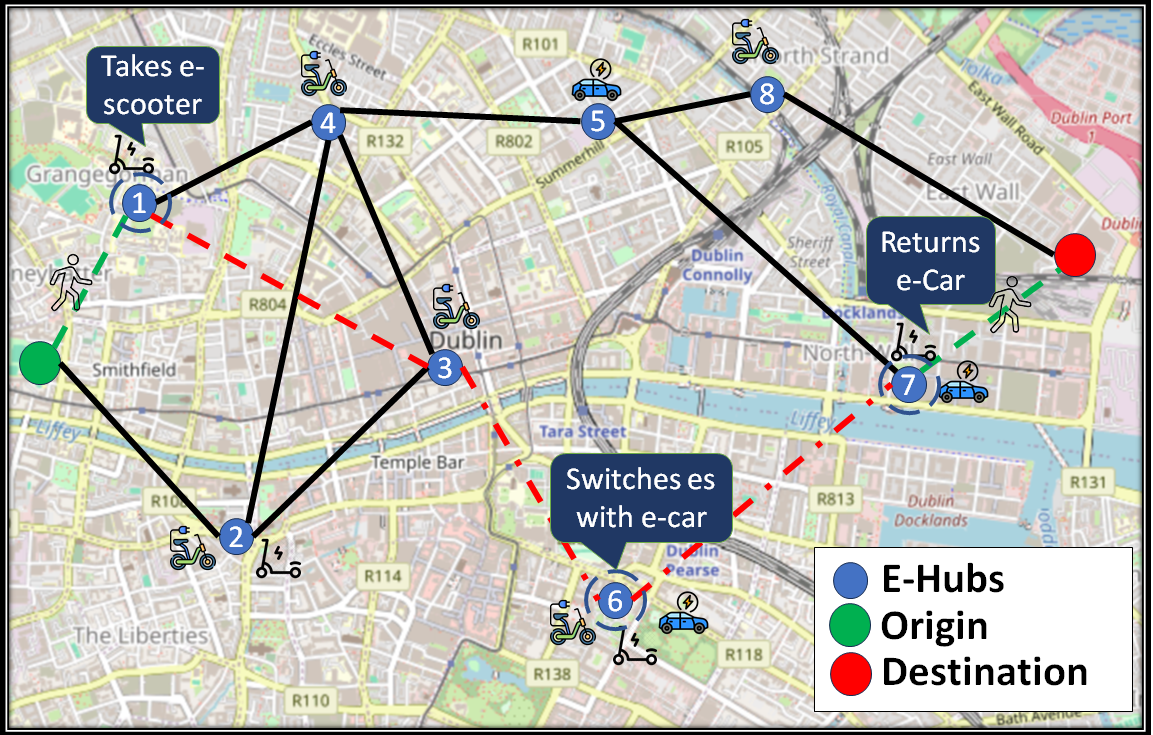}
\caption{Scenario illustration for placement and distribution of e-hubs and multi-modal options}
\label{fig:illustration}
\end{figure}


\section{Problem Description and Formulation}
\subsection{Problem Description}
We present a shared e-mobility framework designed to assist users in planning optimal multi-modal routes based on their preferences and practical constraints. Fig. \ref{fig:illustration} shows a scenario in which e-hubs are deployed in an urban environment, offering various e-mobility options. The user’s journey, from point ``O" (green flag) to point ``D" (red flag), involves navigating through these hubs. Walking is always available, while users can specify preferences, such as avoiding certain modes or limiting the number of transitions. For example, a user who wishes to avoid e-scooters can rely on the algorithm to find a route without them. The algorithm also adapts to constraints like the maximum allowable transitions between modes.
For default preferences, the framework evaluates all available options to suggest the most time-efficient route, potentially combining walking, e-scooters, e-bikes, and e-cars. Mode transitions are allowed only at e-hubs that support both incoming and outgoing modes, ensuring practical viability. For instance, a user following the path in Fig. \ref{fig:illustration} walks to node 1, switches to an e-scooter at node 6, then transitions to an e-car at node 7, and walks the final segment to the destination. 
To address range anxiety, we incorporate energy constraints based on the state of charge (SOC) and energy consumption rates of each mode. Additionally, starting nodes only allow outgoing transitions, destinations only incoming ones, and intermediate nodes balance transitions for smooth operation.

\subsection{Notations}
To formulate the problem, we first define some notation that is subsequently used in this paper. Let \( \mathcal{G} = (\mathcal{V, E, W}) \) be a weighted directed graph where \( \mathcal{V} = \{v_1, \ldots, v_n\} \) represents road segments or nodes, and \( \mathcal{E} = \{e_1, \ldots, e_m\} \) is the set of edges with associated weights \( \mathcal{W}^{(s)} = \{w_1^{(s)}, \ldots, w_m^{(s)}\} \). 
The weight \( w_i^{(s)} \) denotes the travel time cost associated with using mobility mode \( s \) on edge \( e_i \), where \( s \in \mathcal{M} \) and $\mathcal{M} = \{\text{e-car}, \text{e-bike}, \text{e-scooter}, \text{walk}\}$
includes the full set of mobility modes. The adjacency matrix \( \mathcal{A} \) is of size \( n \times n \), where \( \mathcal{A}_{ij} = 0 \) if there is no edge from vertex \( i \) to vertex \( j \), and \( \mathcal{A}_{ij} = 1 \) otherwise. The time cost of transitioning between nodes using a mode \( s \) is denoted by $\mathcal{C}_{v_iv_j}^s$, where \( v_j \in \mathcal{N}(v_i) \) indicates the neighbors of \( v_i \). The time cost associated with changing mode from \( s \) to \( s' \) within node \( v_i \) is represented by \( \mathcal{D}_{v_i}^{ss'} \). We also define binary decision variables \( \mathcal{X}_{v_iv_{j}}^s \) and \( \mathcal{Y}_{v_i}^{ss'} \). The variable \( \mathcal{X}_{v_iv_{j}}^s \) indicates whether the transition from node \( v_i \) to \( v_j \) utilizes mobility mode type \( s \), while \( \mathcal{Y}_{v_i}^{ss'} \) represents the mode transition within node \( v_i \) from e-mobility type \( s \) to \( s' \). It is important to note that transitions between different nodes must consider the same \( s \), as this ensures that an e-mobility tool is returned to a station of the same type. For example, an e-car picked up at node \( v_i \) cannot be returned to an e-bike station at \( v_j \). Given this context, our optimization problem is to find a path \( p = \langle v_1^s, v_2^s, \dots, v_h^s \rangle \) which consists of a sequence of vertices and corresponding mode (\( s \)) starting from the source vertex \( v_1 \in \mathcal{V} \) and terminating at the destination vertex \( v_h \in \mathcal{V} \) that minimizes a defined travel cost, detailed in the following section. Essentially, we aim to determine the optimal multi-modal options, within a set of constraints, that enable the user to navigate each road segment along the path. We shall formulate this problem as a constrained optimization problem and apply commercial and open-source MILP solvers as well as modified Dijkstra's algorithm to find the optimal solution. In the following, we provide a detailed formulation.

\subsection{Problem Formulation}
Based on the key notations mentioned above, we now define the following objective function:

\begin{equation}
\quad \sum_{i=1}^{h-1} \sum_{v_j \in N(v_i)} \sum_{s \in \mathcal{M}} \mathcal{C}_{v_iv_j}^s \mathcal{X}_{v_iv_j}^s \\
+ \sum_{i=1}^{h}  \sum_{s, s' \in \mathcal{M}, s \neq s'} \mathcal{D}_{v_i}^{ss'} \mathcal{Y}_{v_i}^{ss'}.
\label{eq:objective}
\end{equation}


The objective function (\ref{eq:objective}) encapsulates the overall travel time cost corresponding to different e-mobility options \( s \), which is controlled by the decision variable \(\mathcal{X}\). Similarly, the cost of transition between different e-mobility options at a certain node is controlled by the decision variable \(\mathcal{Y}\). To ensure that the travel time cost \( \mathcal{C}_{v_iv_j}^s \) in the objective function accurately reflects real-world conditions, we define it as:

\begin{equation}\label{eq:cost}
\mathcal{C}_{v_iv_j}^s = \alpha_s \cdot \frac{d_{v_iv_j}}{\textit{V}_{v_iv_j}^s} \in \mathcal{W}^{(s)}
\end{equation}

\noindent where the distance between \(v_i\) and \(v_j\) is denoted by \(d_{v_iv_j}\) in metres, and \(V_{v_iv_j}^s\) is the average speed corresponding to the mobility option \(s\) in (m/s). \(\alpha_s\) denotes the preference factor, capturing user preferences for specific e-mobility options. By default, \(\alpha_s\) is set to 1 for preferred modes and is assigned a significantly higher value, e.g., $10^5$, for non-preferred options. For example, if a user wishes to exclude e-cars from recommended options, the corresponding \(\alpha_s\) should be set to a very large value. It is important to note that the average speeds on corresponding edges for each e-mobility mode are determined based on various traffic patterns simulated using the Simulation of Urban Mobility (SUMO) tool, as implemented on the platform \cite{shah2024optimal}. These simulations utilize actual traffic census data to closely replicate real-world conditions. Here, average speeds should be interpreted as snapshot values, calculated at each discrete-time simulation step. To ensure that exactly one mode \(s\) is selected for direct transition between \(v_i\) and \(v_{j}\), we define the constraint below:
\begin{equation}\label{eq:constraint1} 
\sum_{s \in \mathcal{M}} \mathcal{X}_{v_iv_{j}}^s \mathcal{A}_{v_iv_{j}} = 1, \quad \forall i \in \{1, \ldots, h-1\}, \quad v_j \in \mathcal{N}(v_i).
\end{equation}



Another crucial aspect is to ensure flow conservation, which maintains a balance in transportation modes entering and leaving each vertex in the network. Moreover, it is important to ensure that target nodes support the common mode of transport to dock the previous tool for any mode transition (from \( s \) to \( s' \)), thereby assuring smooth mode transitions within the transportation network based on actual compatibility. This aspect is formulated as the following constraint:


\begin{equation}\label{eq:constraint3}
\begin{aligned}
\sum_{v_j \in \mathcal{N}(v_i)} \mathcal{X}_{v_j v_i}^s 
&= 0, 
\quad \forall s \in \mathcal{M}, \quad v_i = v_1, \\
\sum_{v_j \in \mathcal{N}(v_i)} \mathcal{X}_{v_i v_j}^s 
&= 0, 
\quad \forall s \in \mathcal{M}, \quad v_i = v_h, \\
\sum_{v_j \in \mathcal{N}(v_i), v_j \neq v_h} \mathcal{X}_{v_i v_j}^s 
&= \sum_{v_k \in \mathcal{N}(v_j)} \mathcal{X}_{v_j v_k}^s 
+ \mathcal{Y}_{v_j}^{ss'}, \\
& \forall v_i \in \mathcal{V}, \forall s, s' \in \mathcal{M}, s \neq s'.
\end{aligned}
\end{equation}

Since we have formulated the objective function based on binary decision variables, we have to ensure that these variables are strictly binary values, using the following constraint:

\begin{equation}\label{eq:constraint4} 
    \mathcal{X}_{v_iv_{j}}^s, \mathcal{Y}_{v_i}^{ss'} \in \{0, 1\} 
\end{equation}

To incorporate the user preference pertaining to the maximum number of transitions between the e-mobility tools we enforce the following constraint, where, \( T_{\text{max}} \) represent the maximum number of allowed transitions between different e-mobility tools.

\begin{equation}\label{eq:constraint5}
    \sum_{i=1}^{h} \sum_{s, s' \in \mathcal{M}, s \neq s'} \mathcal{Y}_{v_i}^{ss'} \leq T_{\text{max}}
\end{equation}

Finally, an important aspect to consider is energy. To do this, we incorporate the energy consumption constraints which track the energy level required for the next move by the respective e-mobility tool to make the recommendation for optimal route. Considering \(E_s^{v_i}\) to be the initial energy available for the mode $s$ in (Wh) at node \(v_i\), and \(\rho_s\) to be the energy consumption rate per unit distance for the mode \(s\). We note that each e-mobility type \(s\) may include multiple e-mobility tools available at the hub. For simplicity, we assume \(E_s^{v_i}\) represents the vehicle with the highest SOC for mode \(s\) at node \(v_i\). The energy constraint for move from \(v_i\) to \(v_j\) can be represented by:
  
\begin{equation}\label{eq:constraint6}
E_s^{v_j} = E_s^{v_i} - \rho_s \cdot d_{v_i v_j}, \quad E_s^{v_j} \geq 0, \quad \forall s \in \mathcal{M} \setminus \{\text{walk}\}.
\end{equation}

Therefore, our optimization framework seeks to minimize the objective function in (\ref{eq:objective}), subject to the constraints outlined in (\ref{eq:cost}) - (\ref{eq:constraint6}). 

\section{Algorithms}
In this section, we provide a brief overview of two different approaches that we employ to solve the above mentioned constrained optimization problem. 
\begin{figure*}[t]
\centering
\includegraphics[width=\textwidth, height = 6cm]{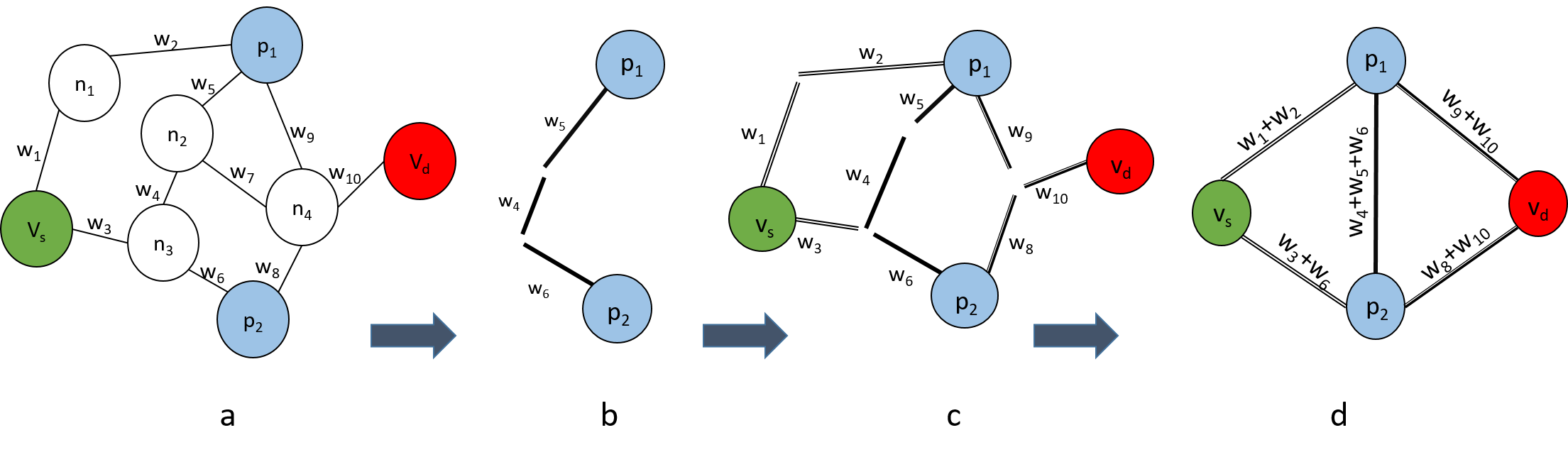}
 \caption{Illustration of the graph reduction methodology. a) Shows the sample graph with (k=2) e-hubs, b) Shows the shortest path between e-hub pairs c) shows the shortest path between origin and destination to e-hubs d) Illustrates the final reduced graph}
 \label{fig:illustration_graph}
\end{figure*}

\subsection{Mixed Integer Linear Programming (MILP)}
To solve the shared e-mobility problem, one approach is to use MILP solvers, which offers significant flexibility due to its inherent ability to model complex constraints and handle multiple objectives. MILP solvers can effectively minimize the given objective function, as defined in eq.(\ref{eq:cost}), using both open-source and commercial solvers such as Gurobi \cite{gurobi}, CBC through PuLP \cite{PulpCBC}, and GLPK \cite{GLPK}. Each of these solvers employ distinct techniques such as branch-and-bound and branch-and-cut tailored for MILP problems.
However, a major drawback of using MILP is its computational complexity. This complexity grows exponentially with the problem size, specifically, the number of decision variables and constraints, which is directly influenced by the graph size in shared mobility networks. Large graphs with numerous nodes, edges, and constraints make MILP solvers more prone to higher memory usage and longer computation times \cite{complexity_SP}. Additionally, as constraints become more intricate (e.g., those related to time windows, capacity, and preference nodes), the solver may require advanced techniques like cutting planes or heuristics to converge within reasonable time limits.
To address this challenge, we propose a graph reduction approach tailored to the shared e-mobility context. By strategically reducing the graph such as aggregating non e-hub nodes we can significantly decrease the problem size without sacrificing solution quality for certain scenarios.  We will elaborate on this graph reduction technique and its limitation in the following:
\subsubsection{Graph Contraction for MILP based Shared E-mobility Solution}
The proposed graph contraction method, inspired by contraction hierarchies used in road network routing \cite{contractionhierarchies}, offers a scalable solution tailored to the shared e-mobility framework described previously. It leverages e-hubs associated with deployed e-mobility vehicles to optimize the network efficiently. 
The primary objective is to retain the nodes supporting e-hubs with their associated weight information while eliminating redundant nodes from the graph network. To make the process more intuitive for readers, consider the illustration in Fig. \ref{fig:illustration_graph}, using a small sample graph with (\(k=2\)) e-hubs. The goal is to reduce the network to (\(k+2\)) nodes, which will then be used to implement the constrained user-centric optimization framework formulated in section II.
In the pre-computation step, we find the shortest paths between all e-hub pairs. For instance, in the illustration with two preferred stations (\(p_1\) and \(p_2\)), if the shortest path is \(p_1 \rightarrow n_2 \rightarrow n_3 \rightarrow p_2\), we retain \(p_1\) and \(p_2\) with the cumulative weight \(w5 + w4 + w6\), as shown in Fig. \ref{fig:illustration_graph} (b).
Next, we determine the unconstrained shortest routes between the start node \(v_s\), the end node \(v_d\), and all preferred nodes (\(p_1\) and \(p_2\)). Specifically, we find the shortest paths for \(v_s \rightarrow p1\), \(v_s \rightarrow p2\), \(v_d \rightarrow p1\), and \(v_d \rightarrow p2\) as shown in Fig. \ref{fig:illustration_graph} (c). We retain the weight information on these edges and contract the nodes to generate a new graph with \(k+2\) nodes, as shown in Fig. \ref{fig:illustration_graph} (d).
While this illustrative example is just to provide an intuitive picture, in practical scenarios, such as the one we consider for validation, the overall number of nodes in the graph are significantly large as compared to the  e-hubs, thereby providing a substantial practical advantage in terms of run-time complexity. The next section details the experimental evaluation of the proposed framework for optimal cost, adaptability to varying scenarios, and computational efficiency to demonstrating its effectiveness. 
\subsubsection{Limitations of Graph Reduction Approach} The proposed approach is a heuristic method and does not offer a universally optimal solution across diverse problem settings. However, in scenarios where the number of e-hubs is limited, it demonstrates significant computational benefits in conjunction with MILP-based solutions. By focusing on the shared e-mobility context, this reduction method helps streamline the problem by selectively pruning irrelevant nodes and edges, resulting in reduced problem dimensions. This enables solvers to find near-optimal solutions more efficiently, with notable improvements in execution time and scalability when dealing with smaller, more manageable graphs.
The detailed performance evaluation based on the proposed framework with and without the application of reduced graph approach is presented in Section \ref{sec:eval} for different number of e-hubs. We note that this paper emphasizes providing a practical framework for shared e-mobility, and as such, we do not delve into the formal mathematical bounds. Future work will focus on refining this method to enhance its applicability across different map sizes and graph topologies. This will involve developing more sophisticated graph pruning strategies and potentially incorporating dynamic updates as the problem scales, enabling broader applicability and stronger theoretical guarantees.

\subsection{Modified Multi-modal Constrained Dijkstra's Algorithm}
Dijkstra's algorithm is a well-established method for finding the shortest path, guaranteeing an optimal solution for single-source shortest path problems \cite{dijkstra_original}. However, the classic version is not designed to handle multi-modal objective with constraints. To align it with our problem requirements, we adapted Dijkstra's algorithm to account for multi-modal options, user preferences and constraints, such as SOC, mode transitions and ensuring common modes on transition e-hubs. Although this modification introduces flexibility, it does not always yield the global optimum due to the complexity introduced by the constraints.
Our adaptation introduces multi-modality by extending the NetworkX graph definition. We add attributes representing virtual edges at nodes corresponding to e-hubs where different transportation modes (e.g., e-car, e-bike, e-scooter) are available. This setup allows us to enforce additional constraints, including common mode availability, energy limitations, and the maximum number of allowed mode transitions.
For implementation we use Python’s `heapq' library \cite{heapq} to efficiently manage a priority queue for node exploration. The algorithm initializes the queue with the starting node, setting the initial mode to ``walk" alongside the SOC and other relevant parameters. As the algorithm processes each node, it ensures that transitions comply with energy and mode constraints, updating SOC and mode availability as needed. User preferences are incorporated by prioritizing transitions that align with preferred modes and minimizing unnecessary transitions between modes.
This modification maintains the priority queue's efficiency while handling real-world constraints, ensuring a practical and computationally feasible solution with significantly better computational complexity as compared to the MILP approach. The complete pseudo-code of the modified algorithm is provided in the Algorithm (\ref{algo:dijkstra_energy}).
\begin{algorithm}[htb]
\caption{Modified Multi-Modal Dijkstra's Algorithm with Constraints}
\label{algo:dijkstra_energy}

\KwIn{Graph \(\mathcal{G} = (\mathcal{V}, \mathcal{E}, \mathcal{W})\), start node \(v_s\), end node \(v_d\), ,user preferences, e-hub locations}
\KwOut{Optimal path from \(v_s\) to \(v_d\) with minimal travel time with constraints}

\textbf{Initialization} \;
Priority queue \texttt{pq} with tuple \((0, v_s, \text{walking}, 0, \text{$E_s^{v_i}$})\) \;
Distances: \(d[v] = \infty\) $\forall$ \(v \in \mathcal{V}\), set \(d[v_s] = 0\)\;
Transitions count: \(\text{transitions}(v) = 0\) $\forall$ \(v \in \mathcal{V}\) \;
Initial Energy (SOC): \(\text{SOC}_s(v) = \text{SOC}_s^0\) $\forall$ \(v \in \mathcal{V}\) and all modes \(s\) \;
Empty set \texttt{visited} to store visited nodes\;

\textbf{Main Loop} \;
\While{\texttt{pq} is not empty}{
Select and remove the entry with the smallest time value from \texttt{pq}, retrieving its associated node, mode, transitions, SOC, and path \;

    \If{\(\text{current\_node} = v_d\)}{
        \Return optimal path \text{path} and \text{current\_time} \;
    }
    
    \If{\((\text{current\_node}) \in \texttt{visited}\)}{
        \textbf{continue} \;
    }
    
    Mark \((\text{current\_node})\) as visited \;

    \For{each neighbor \(v\) of \text{current\_node}}{
        \For{each mode \(s\) available at \(v\)}{
            \If{mode transition from \text{current\_mode} to \(s\) is allowed}{
                Calculate travel time and energy consumption for transition \;
                \If { \(\text{available energy in mode }  \text{at current node}\newline \geq \text{energy required for the transition}\)}{
    Update the energy state for mode \(s\) \;
    Add the next node, mode, updated travel time, and path information to the priority queue \;
}

                }
            }
        }
    }
\Return \text{``No feasible path found"} \;

\end{algorithm}








\section{Evaluation}  \label{sec:eval}
\subsection{Experimental Setup}

The overall scenario is managed through the platform \cite{shah2024optimal}, which leverages SUMO to create diverse traffic scenarios for the Dublin City Centre (DCC) map. Real-time traffic updates from the platform are used to calculate average speeds, which are then incorporated into equation (\ref{eq:cost}) to reflect realistic conditions. The real-time speed data is fetched dynamically from the cloud as edge weights to be incorporated in a weighted, directed graph \((\mathcal{G})\), generated using NetworkX \cite{networkx_hagberg2008exploring} library, ensuring the model adapts to changing traffic patterns. The number and locations of e-hubs are also managed by the backend simulator, with experiments conducted using 10, 20, 50,  100, 200 and 300 e-hubs on the same map to evaluate the proposed framework using MILP solvers with graph contraction approach and also for modified Dijkstra's algorithm. We assign all e-mobility options uniformly on the e-hubs for these experiments to focus on the algorithm's adaptability for user-preference and multi-modality.  
In essence, the evaluation is focused on: (1) Assessing the computational complexity and optimal time cost of MILP solution with and without graph reduction method for different number of e-hubs (2) Computational complexity and optimal time cost performance of modified Dijkstra's algorithm (3) Adaptability of both algorithms to different user-preferences and constraints. 
For statistical evaluation we use 500 (O, D) pairs with varying distances on the map and are consistently used for all the subsequent experiments. 
\subsection{MILP Evaluation with \& without Graph Reduction}
\begin{figure}[!htb]
\centering
\includegraphics[width=\linewidth, height = 6 cm ]{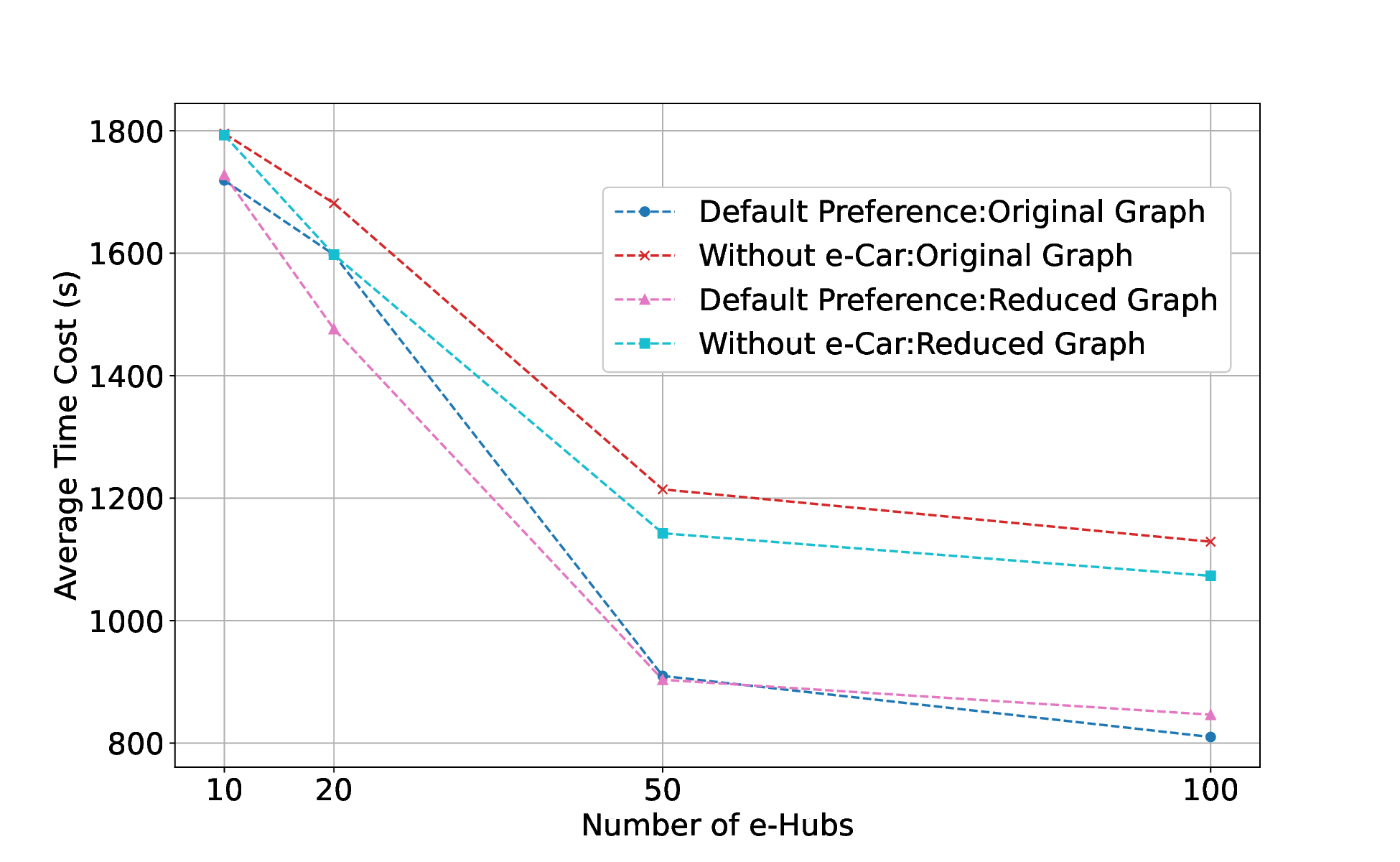}
 \caption{Objective cost comparison between Original and Reduced Graphs for the Gurobi Solver under default user preference and scenario where e-cars are not preferred}
 \label{fig:ObjectivCost}
\end{figure}

To evaluate the performance of the MILP solution, we used a consistent set of 500 origin-destination (O-D) pairs on the Dublin City Center (DCC) map, testing two sets of user preference: (1) the default option, which includes all modes, and (2) a restricted preference excluding electric cars. Both original and reduced graph methods were applied to these preferences. The results, illustrated in Fig. \ref{fig:ObjectivCost}, show that the Gurobi solver consistently performs well under both preference sets, though the default preferences (all modes) yield lower time costs compared to the case without e-cars. The reduced graph method generally provides lower average time costs, particularly when e-cars are excluded. However, for default preferences, the cost difference between the original and reduced graphs is minimal. Moreover, scenarios with fewer e-hubs typically result in higher average time costs, as the algorithm defaults more frequently to walking. The time cost difference between using 50 and 100 e-hubs is marginal, prompting the focus on 50 and 20 e-hubs to assess runtime complexity across solvers.

\begin{figure}[!ht]
\centering
\includegraphics[width=0.9\linewidth, height = 6 cm ]{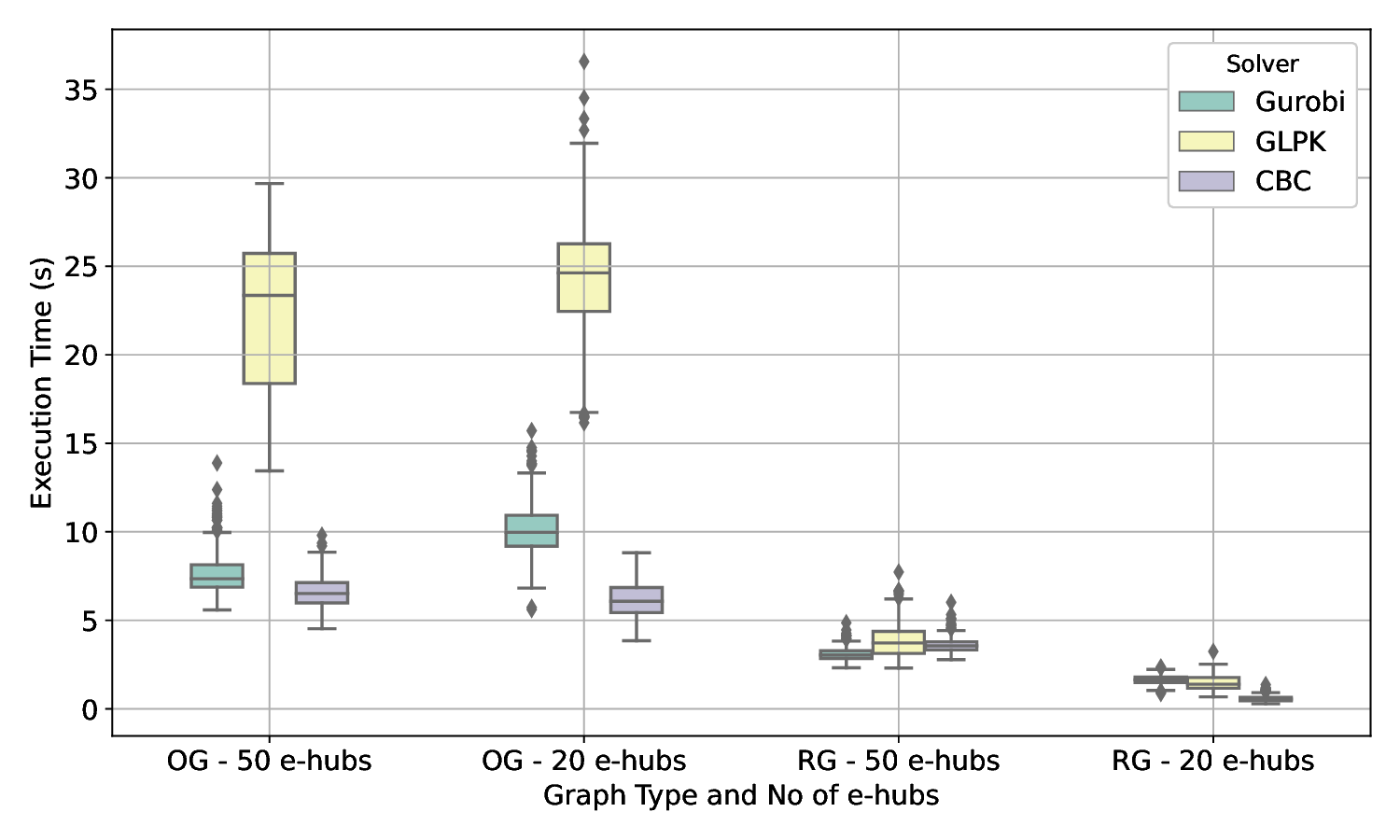}
 \caption{Box plot visualization depicting the distribution of execution times for Original Graph (OG) and Reduced Graph (RG) using 50 and 20 e-hubs across three solvers}
 \label{fig:exectimesvsk}
\end{figure}

\begin{figure}[htb]
\centering
\includegraphics[width=0.5\textwidth, height = 6cm]{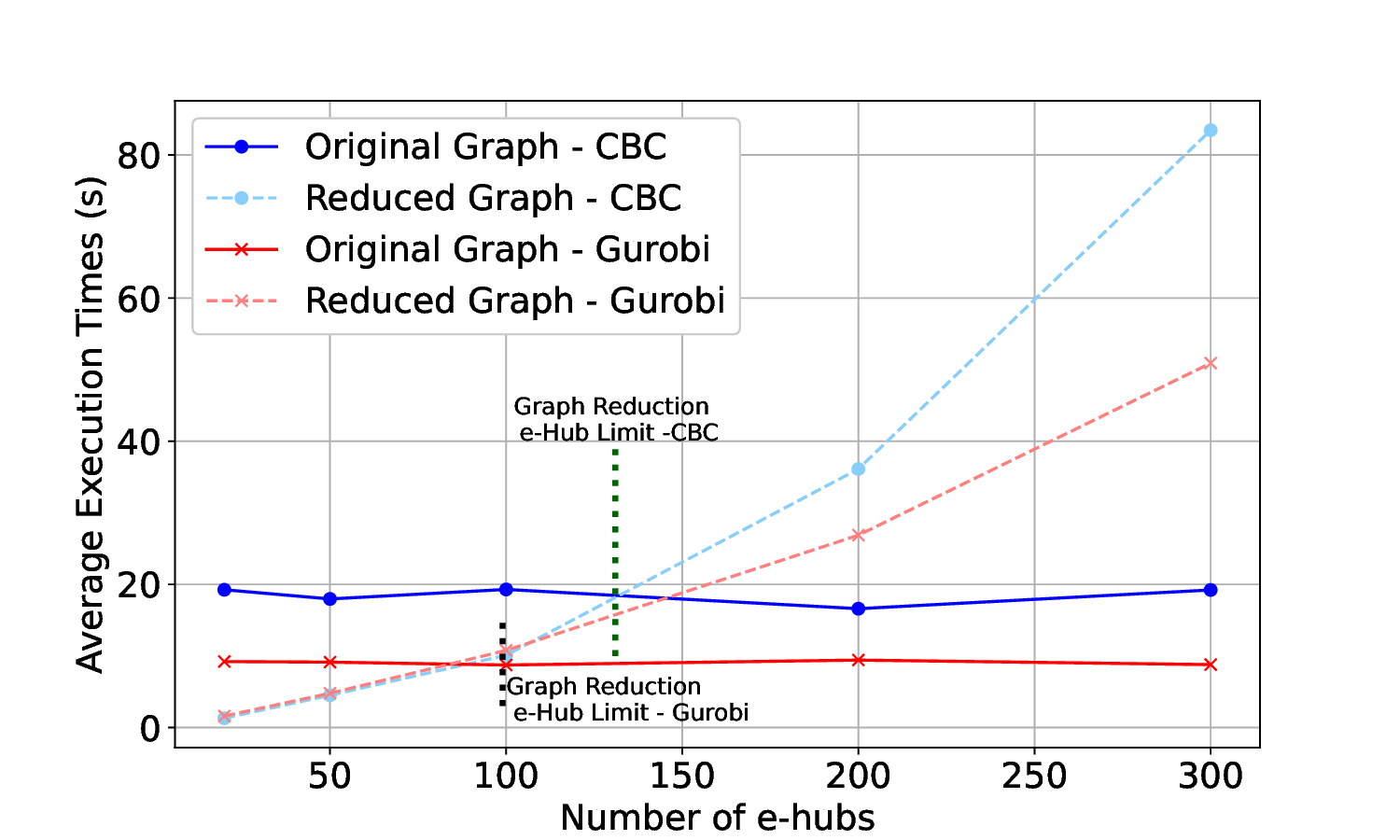}
 \caption{Mean execution times for original and reduced graphs with CBC and Gurobi solvers, evaluated for varying numbers of e-hubs. Thick vertical dotted lines indicate the e-hub limits beyond which graph reduction is no longer effective for each solver.}
 \label{fig:origvsred_meanexectime}
\end{figure}

\begin{figure}[htb]
\centering
\includegraphics[width=0.5\textwidth, height = 6cm]{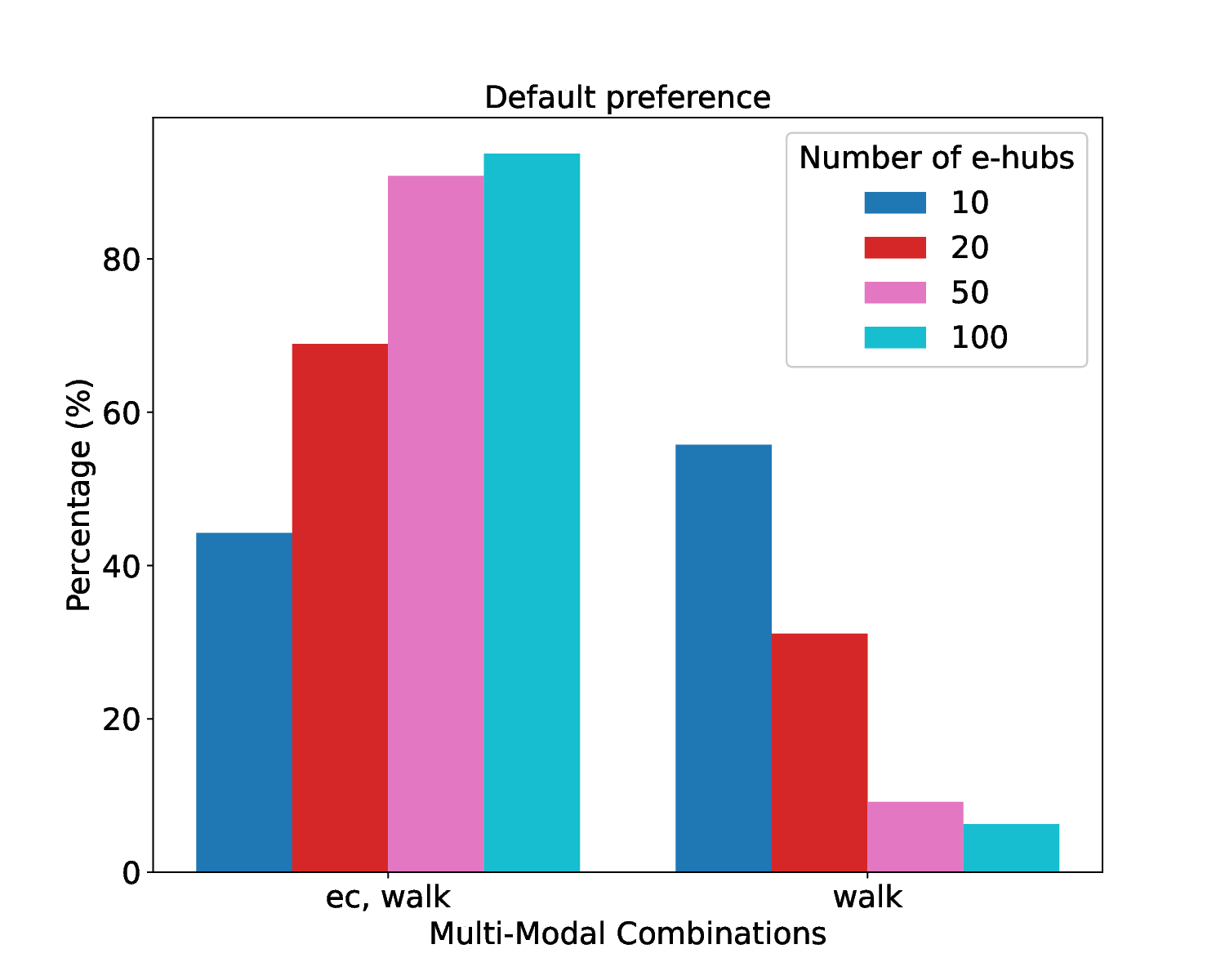}
 \caption{Distribution of multi-modal transportation combinations for varying number of e-hubs, analyzed based on default user preferences (with all options) using the Gurobi solver.}
 \label{fig:multimodalstats}
\end{figure}

\begin{figure}[htb]
\centering
\includegraphics[width=0.5\textwidth, height = 6cm]{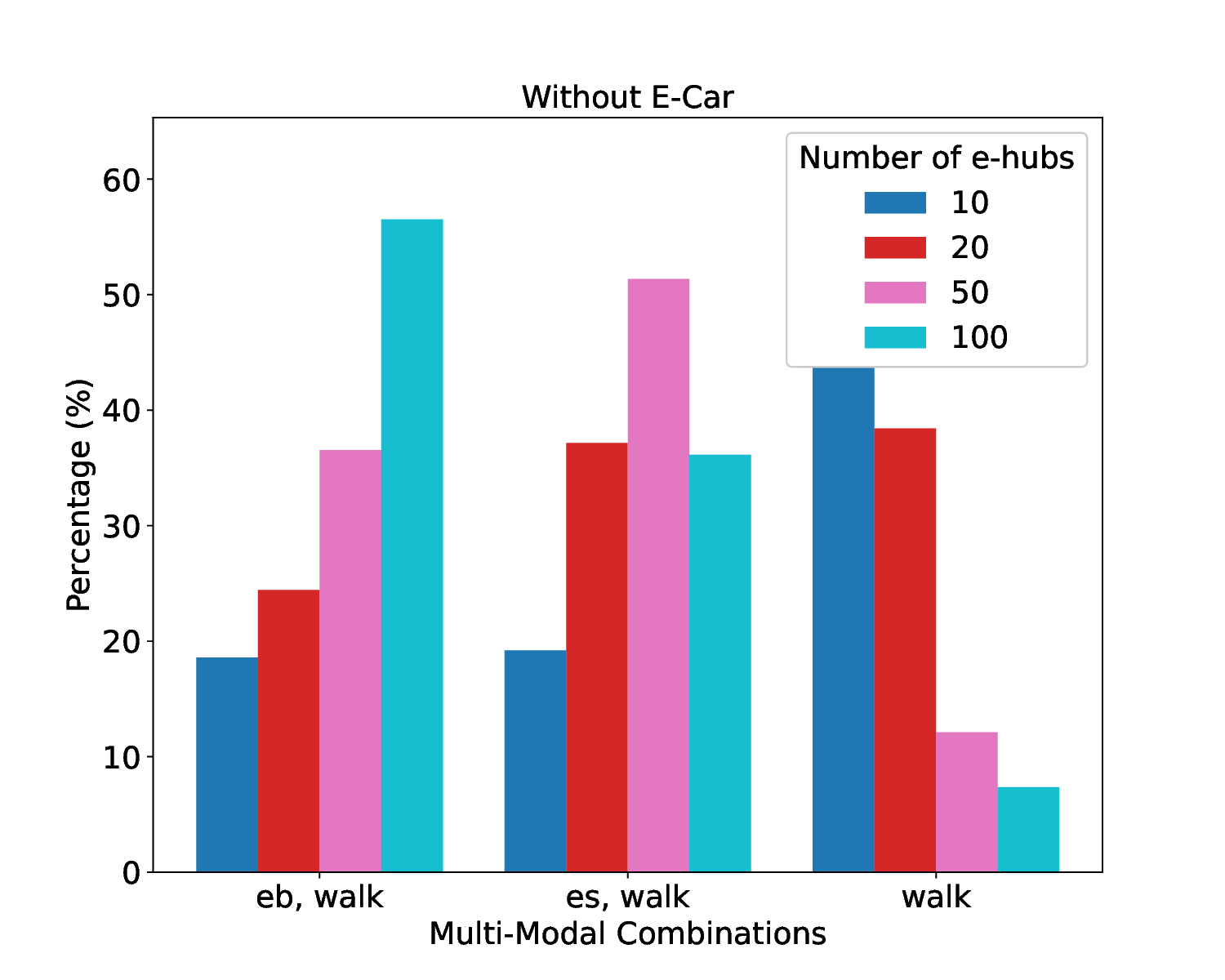}
 \caption{Distribution of multi-modal transportation combinations for varying number of e-hubs, analyzed based on  user preferences without e-car  using the Gurobi solver.}
 \label{fig:multimodalstats_2}
\end{figure}

Fig. \ref{fig:exectimesvsk} highlights the runtime performance for Gurobi, GLPK, and CBC solvers under default preferences, comparing the original and reduced graphs with 20 and 50 e-hubs. For all solvers, the reduced graph significantly reduces median execution time, achieving sub-2 second times for 20 e-hubs across all solvers, with minimal variability. For 50 e-hubs, the median execution time remains efficient, around 3 seconds. In contrast, the original graph exhibits significantly higher runtimes, especially for the GLPK solver, which averages 25 seconds, compared to 7-10 seconds for Gurobi, with outliers extending up to 15 seconds. This represents a run-time efficiency improvement of more than 65\% for 20 e-hubs and 75\% for 50 e-hubs using the reduced graph method. However, for networks with more than 100 e-hubs, the reduced graph method becomes less effective, and the original graph outperforms in terms of scalability.

Fig. \ref{fig:origvsred_meanexectime} further demonstrates the computational advantage of the reduced graph method in terms of the average execution time between varying numbers of e-hubs. For both the Gurobi and the CBC solvers, the reduced graph consistently outperforms the original graph for up to 100 e-hubs. The execution time for the original graph remains largely unaffected by the number of e-hubs, while the reduced graph offers clear runtime benefits for networks with small number of e-hubs, aligning well with micro-mobility scenarios. However, as previously discussed in the limitations section, the reduced graph method is not universally applicable and becomes counterproductive in case of more than 100 e-hubs. Nevertheless, in micromobility contexts, where the geographical area is smaller and e-hub density is typically lower, the proposed framework provides valuable insights into flexible, scalable optimization methods that address user-centric requirements and real-world constraints.

To assess the multi-modal performance of the reduced graph method, we evaluated the statistical distribution of mode combinations for the two user preference sets (default and excluding e-cars) across 10, 20, 50, and 100 e-hubs using the Gurobi solver. Figs. \ref{fig:multimodalstats} and \ref{fig:multimodalstats_2} illustrate these results. For default preferences with 10 e-hubs, the algorithm selects a combination of e-car and walking 44.3\% of the time, while 55.7\% of trips involve walking only. In contrast, when e-cars are excluded, walking dominates, with 62.2\% of trips relying solely on walking. As the number of e-hubs increases, the use of walking-only routes decreases, validating the algorithm’s adaptability to user preferences and the number of available modes. For larger networks, the combination of e-cars and walking becomes increasingly prevalent. When e-cars are excluded, a similar pattern is observed with e-bikes and e-scooters, whose usage increases as the number of e-hubs grows, with a corresponding decline in walk-only routes.

\subsection{Evaluation of modified Dijkstra's Algorithm}
\begin{figure}[!ht]
\centering
\includegraphics[width=1.1\linewidth, height = 6cm ]{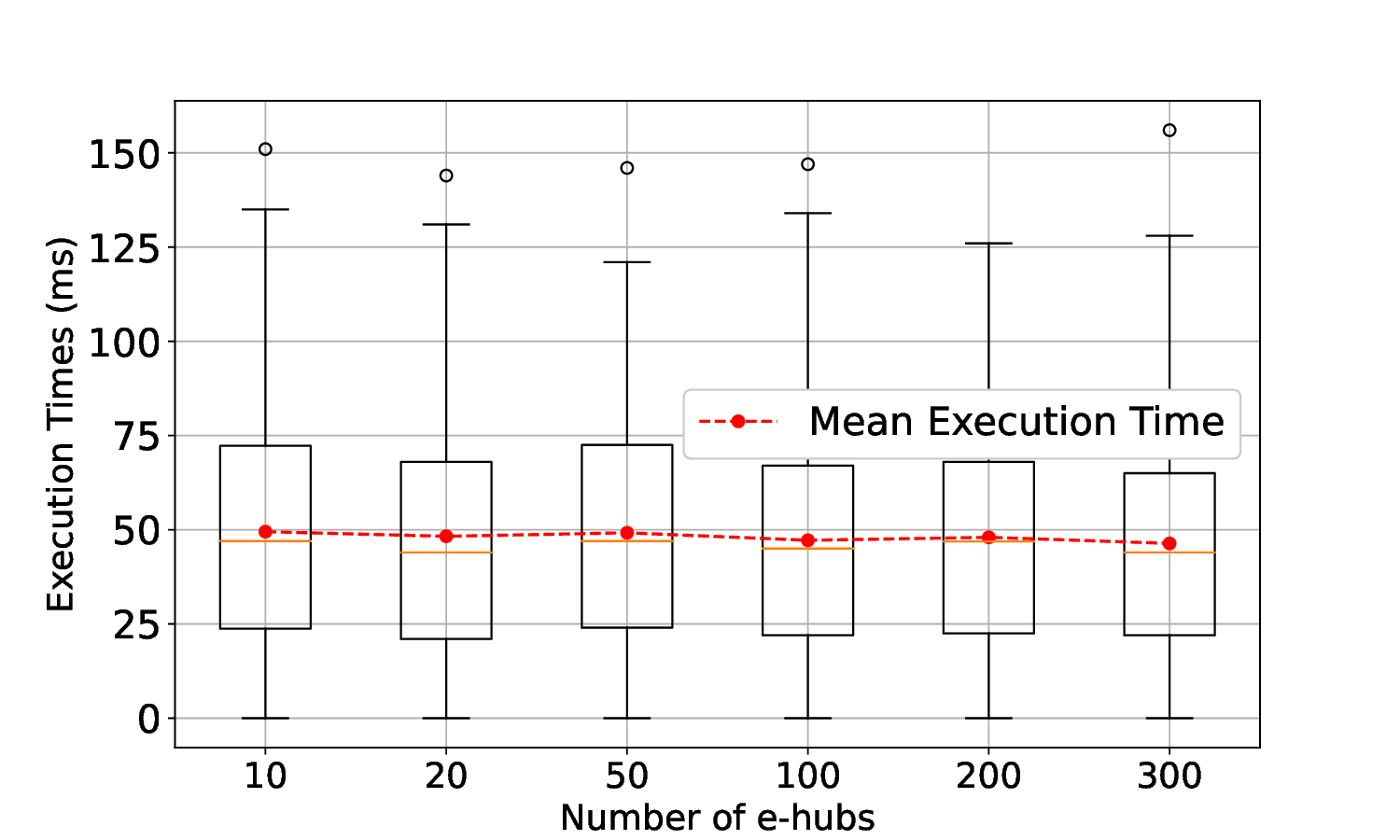}
 \caption{Execution Time Distribution for Dijkstra's algorithm for changing number of e-hubs}
 \label{fig:execution_times}
\end{figure}

\begin{figure}[!ht]
\centering
\includegraphics[width=\linewidth, height = 6cm ]{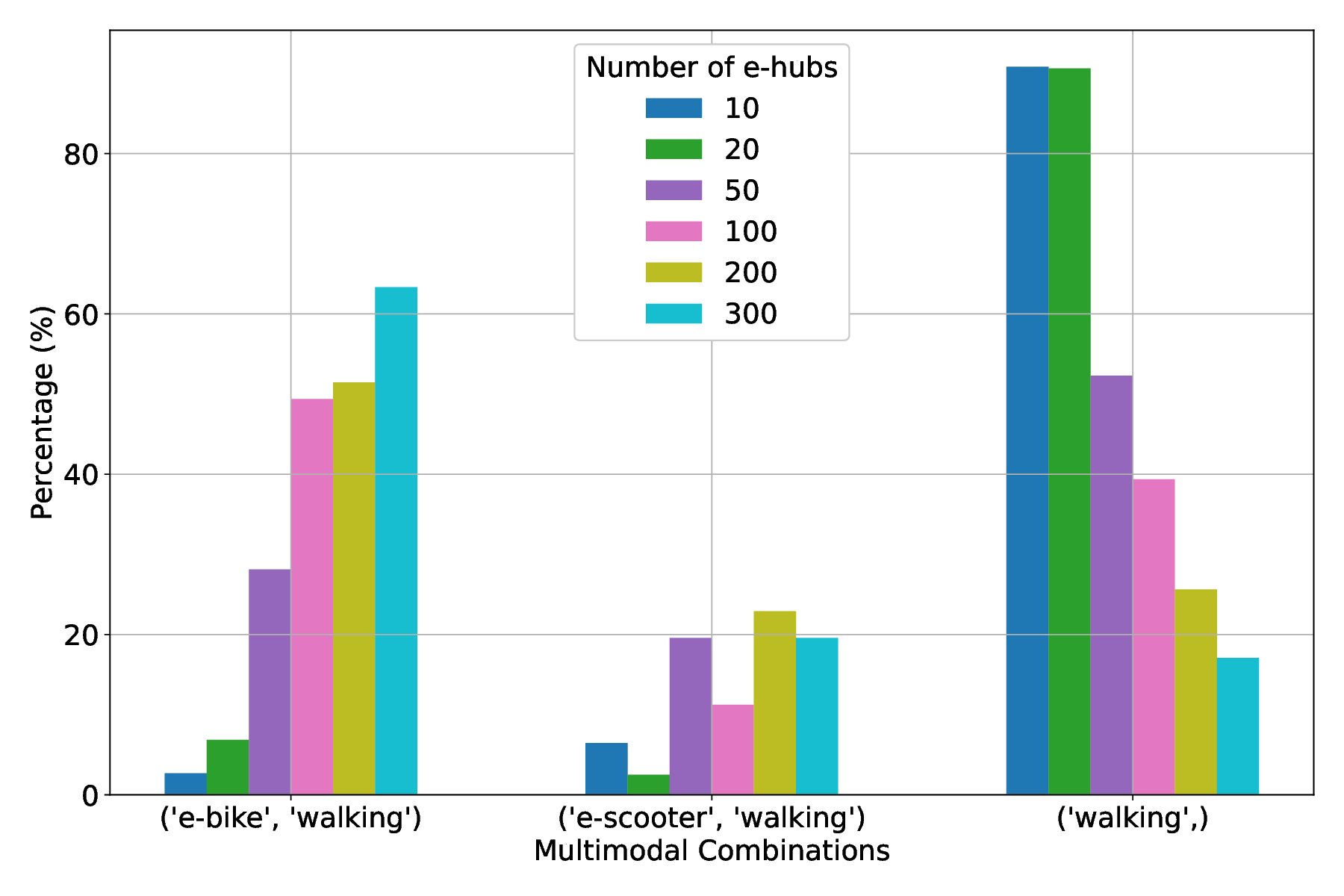}
 \caption{Combination of multi-modal statistics for optimal output when e-car is excluded from the user preference and max allowed transition of 2 using modified Dijkstra's algorithm}
 \label{fig:mmodal_wo_ecar}
\end{figure}

\begin{figure}[!ht]
\centering
\includegraphics[width=\linewidth, height = 6cm ]{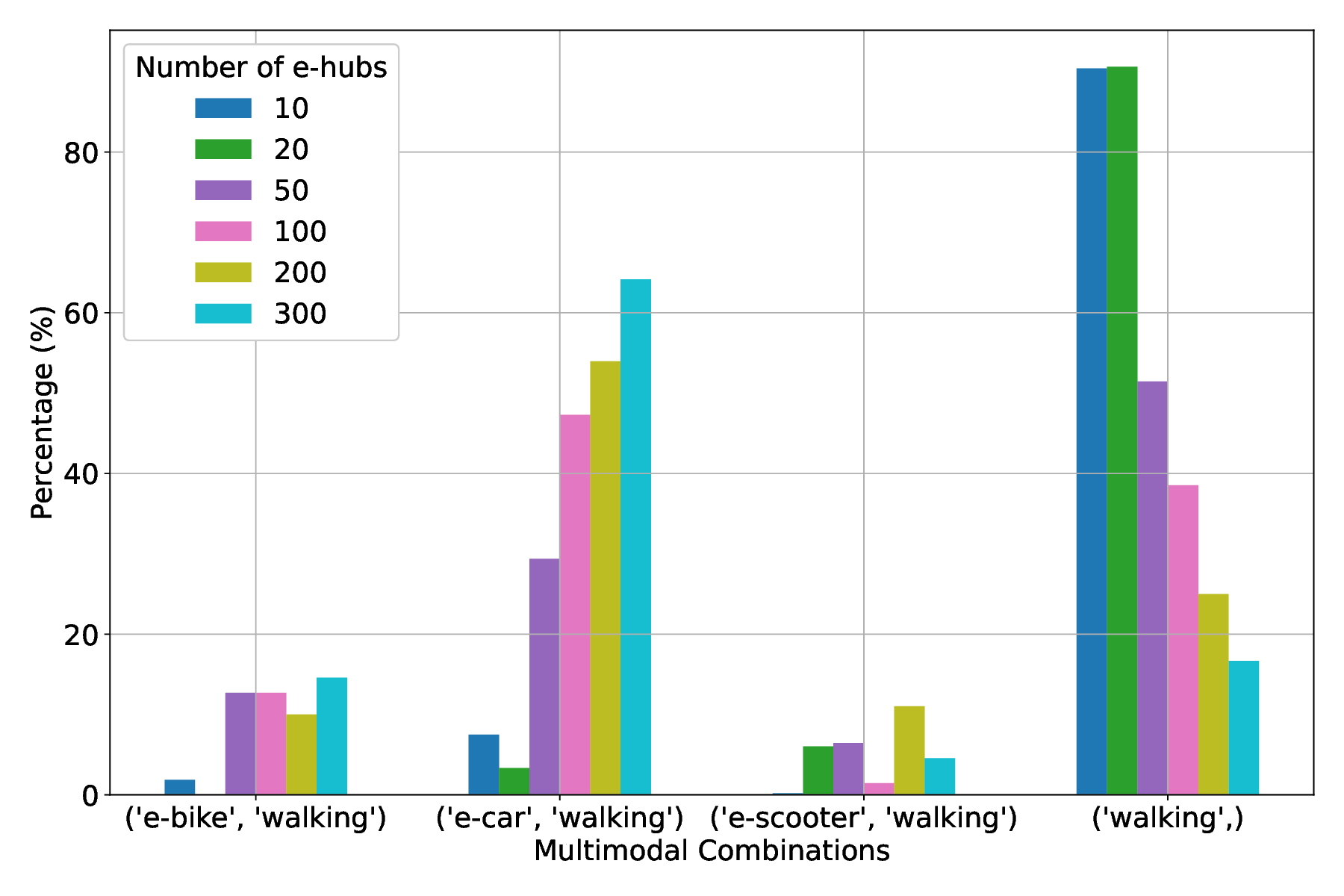}
 \caption{Combination of multi-modal statistics for optimal output with default option including all modes and max allowed transition of 2 using modified Dijkstra's algorithm}
 \label{fig:mmodal_w_ecar}
\end{figure}

To evaluate the modified multi-modal Dijkstra's algorithm, we analyze its runtime complexity and optimal time cost performance against varying user preferences and constraints. Furthermore, we emphasize the algorithm's adaptability by presenting a detailed statistical analysis of its performance across different modalities, highlighting its viability under diverse conditions.
Fig. \ref{fig:execution_times} shows the box plot distribution for execution times of Dijkstra's algorithm with default user preferences and without any energy constraints for varying number of e-hubs. The mean and median execution time doesn't vary significantly for the increasing number of e-hubs with an average execution time of just 50 ms. This is contrary to the graph reduction approach used for the MILP problem in the previous section where the proposed method is only suitable for scenarios where we have less e-hub stations.  
Fig \ref{fig:mmodal_wo_ecar}, and Fig. \ref{fig:mmodal_w_ecar}, provide the statistical outcome for the combination of different modes recommended by the algorithm for the same 500 (O,D) pairs. In Fig. \ref{fig:mmodal_w_ecar}, we use the default user preferences for modes (includes all options) with a maximum allowed iterations of 2, whereas, Fig. \ref{fig:mmodal_wo_ecar} shows the output pertaining to the user-preference where e-car is excluded. The number of e-hubs are varied and the statistics for the combination of output modes are observed across all 500 (O, D) pairs, recommend by the algorithm. It can be seen that the algorithm adapts well to the varying number of e-hubs. For both cases (with and without e-car preferences), the algorithm recommends `walk' only option for more than 85\% of the (O, D) pairs when we have 10 and 20 e-hubs only. The multi-modal combination increases when more e-hubs are deployed. It substantiates that the algorithm adapts well to the user-preferences and multi-modal options are proportional based on the available number of e-hubs.    

\begin{figure}[!ht]
\centering
\includegraphics[width=1.1\linewidth, height = 6cm]{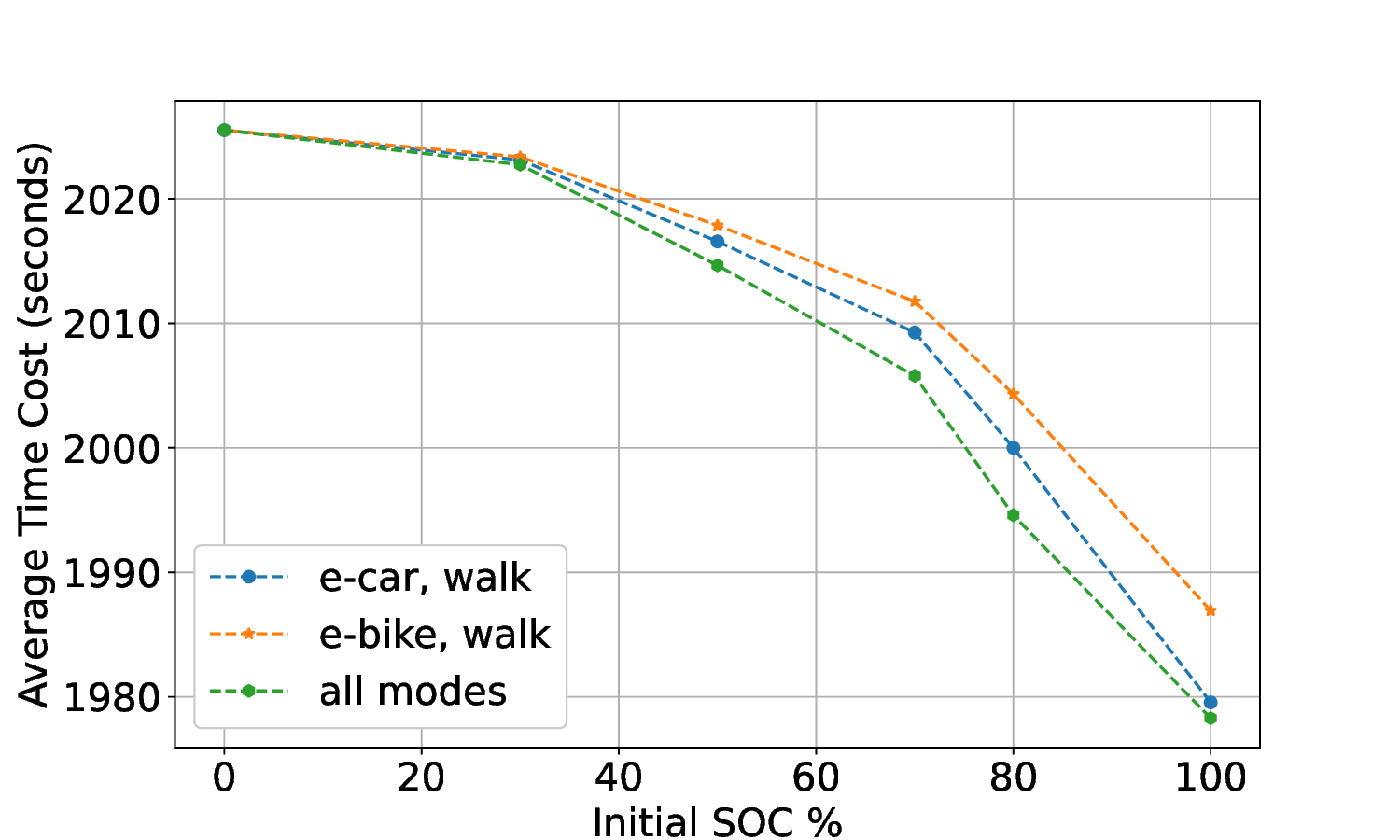}
 \caption{Average Time Cost for different initial battery levels using e-car, e-bike and all modes collectively}
 \label{fig:energy_constratins1}
\end{figure}

\begin{figure}[!ht]
\centering
\includegraphics[width=\linewidth, height = 6cm ]{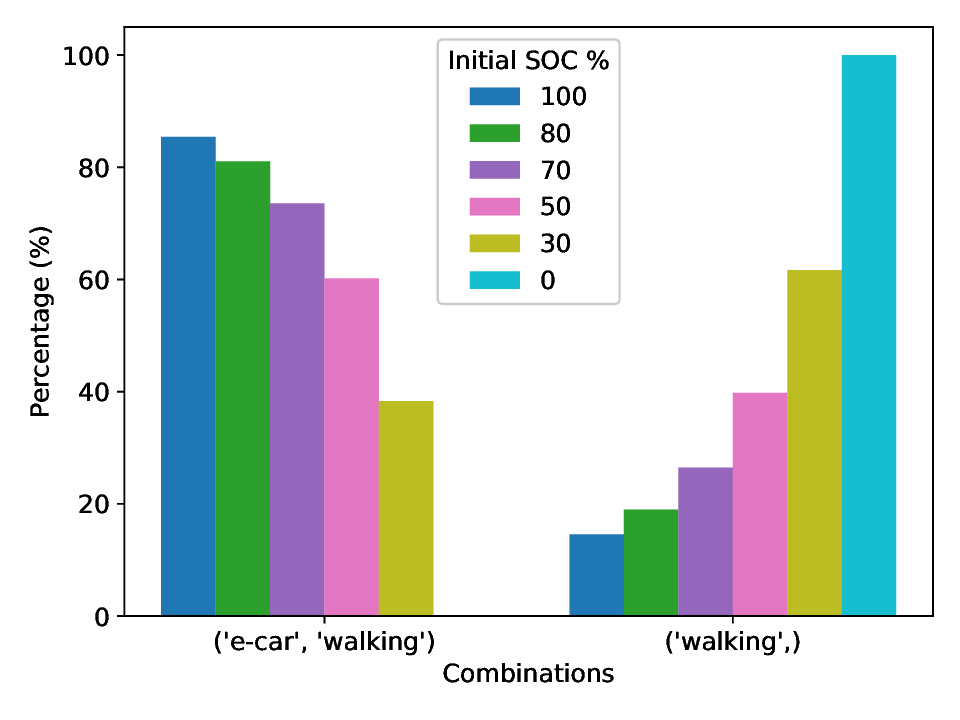}
 \caption{Statistical distribution of mode usage based on initial battery levels for e-car only}
 \label{fig:energy_constratins2}
\end{figure}

Fig. \ref{fig:energy_constratins1} and Fig. \ref{fig:energy_constratins2} show the performance of Dijkstra's algorithm under energy constraints. To evaluate the adaptability of Dijkstra's algorithm with respect to available energy levels, a detailed analysis is provided for individual e-mobility modes. The initial SOC levels are varied for respective e-mobility tools assuming different rate of energy consumption for each e-mobility option. For the sake of simplicity and to focus only on the algorithm's adaptability,  the prevalent energy rates per unit distance are used without taking into consideration the actual energy consumption prediction model \cite{Ding_2024}. Fig. \ref{fig:energy_constratins1} illustrates the average time cost (calculated from 500 origin-destination pairs) for varying initial SOC levels across three distinct e-mobility options. Each curve represents the average optimal time cost under different mobility configurations. Specifically, the blue curve corresponds to the scenario where only e-cars are available alongside the default walking option. The orange curve reflects the case where only e-bikes and walking are available and the green curve depicts the variation in average time cost for different initial SOC levels when all e-mobility modes, e-cars, e-bikes, and walking are accessible. The results demonstrate that the algorithm adapts effectively to each mobility option, both individually and collectively, as the average time cost decreases with increasing initial SOC levels. Furthermore, the average time cost when all modes are available is lower than when only e-cars are available, which in turn is lower than the e-bike-only scenario. This indicates that the algorithm efficiently adjusts to varying SOC levels and optimally leverages the available e-mobility modes based on user preferences. The same outcome is illustrated in Fig. \ref{fig:energy_constratins2}
, which highlights how the statistical distribution of selected modes shifts based on varying initial SOC levels.
\subsection{Discussion}
The results demonstrate that while MILP offers advantages in terms of optimal time cost and flexibility for incorporating multiple constraints, its computational complexity poses a significant challenge for scalable solutions. The proposed reduced graph approach offers a heuristic to mitigate this complexity before applying the MILP solution, but it is not a generalized approach and is effective primarily in scenarios with a limited number of e-hubs.
In real-world applications, the selection and placement of e-hubs depend on factors such as infrastructure, geography, and other considerations. Although the number of e-hubs in this study was chosen based on the map size and aligns with typical deployments \cite{hubs,opt_ehub}, the focus here is on evaluating the performance of MILP and the modified Dijkstra's algorithm within the proposed framework. The MILP evaluation reveals significantly higher computational complexity across all solvers, even when using graph reduction. In contrast, the modified Dijkstra's algorithm offers a more scalable alternative, though MILP remains a viable option due to its flexibility for integrating additional objectives.
Future work will focus on refining and formalizing the graph reduction method to create a more generalized and scalable solution. Additionally, we aim to extend the MILP framework to optimize energy consumption and monetary costs based on user preferences. The proposed framework is open-source, encouraging contributions and enhancements from researchers in this domain.

\section{Conclusion}
In this paper, we introduced a multi-modal optimization framework tailored for shared e-mobility, with a specific focus on micro-mobility settings. We proposed and evaluated two distinct approaches for finding the optimal multi-modal path based on user preferences, while considering practical constraints such as initial SOC, maximum allowable transitions, and the availability of e-mobility options at transfer stations. To address the computational complexity of the MILP approach, we introduced a heuristic-based graph reduction method. This approach mitigates the computational burden but is viable for scenarios with limited number of e-hubs. Additionally, we proposed a modified Dijkstra's algorithm as a more scalable alternative, offering significantly reduced computational complexity while maintaining adaptability to varying user preferences and energy constraints. Both approaches were rigorously evaluated under different scenarios, demonstrating their flexibility in handling diverse constraints. While MILP remains a viable candidate for detailed, constraint-rich optimization problems, the modified Dijkstra's algorithm is better suited for large-scale, real-time applications. Future work will focus on developing scalable graph reduction techniques for complex MILP formulations and integrating metrics such as personalized energy consumption and monetary costs into optimization objectives. We will also investigate multi-objective optimization to balance criteria such as travel time, energy use, and costs, leveraging machine learning-based traffic predictions and addressing e-hub resource uncertainties, building on our prior works in \cite{yan2024u, chen2021comparative}, to enhance robustness and adaptability in real-world shared e-mobility systems.



\section*{Acknowledgment}
This work is supported by Taighde Éireann – Research Ireland, formerly Science Foundation Ireland, under Grant Number \textit{21/FFP-P/10266} and SFI/12/RC/2289\_P2 (Research Ireland Insight Centre for Data Analytics).


\ifCLASSOPTIONcaptionsoff
  \newpage
\fi

\bibliographystyle{IEEEtran}
\bibliography{IEEEexample}

\end{document}